\documentclass[onecolumn, tighten]{aastex631}
\usepackage{nicefrac}
\usepackage{microtype}
\usepackage{booktabs}
\usepackage{rotating}
\usepackage{xcolor}
\usepackage[utf8]{inputenc}
\usepackage{amsmath}

\DeclareUnicodeCharacter{2212}{\ensuremath{-}}

\received{?}
\revised{?}
\accepted{?}
\submitjournal{Frontiers}
\shorttitle{Detection of Planetary Radio Emissions via SKA}
\shortauthors{Bagheri et al.}

\begin{document}

\title{Exploring Radio Emissions from Confirmed Exoplanets Using SKA}

\author{Fatemeh Bagheri}
\affil{NASA Goddard Space Flight Center, Greenbelt, MD, USA}
\affil{Department of Physics, University of Texas at Arlington, Arlington, TX, USA}
\correspondingauthor{Fatemeh Bagheri}
\email{fatemeh.bagheri@nasa.gov}

\author{Anshuman Garga}
\affil{Department of Physics, University of Texas at Arlington, Arlington, TX, USA}

\author{Ramon E. Lopez}
\affil{Department of Physics, University of Texas at Arlington, Arlington, TX, USA}

\begin{abstract}

%%% Leave the Abstract empty if your article does not require one, please see the Summary Table for full details.

\noindent Currently, our understanding of magnetic fields in exoplanets remains limited compared to those within our solar system. Planets with magnetic fields emit radio signals primarily due to the Electron Cyclotron Maser Instability mechanism. In this study, we explore the feasibility of detecting radio emissions from exoplanets using the Square Kilometre Array (SKA) radio telescope. Utilizing data from the NASA Exoplanet Archive, we compile information on confirmed exoplanets and estimate their radio emissions using the RBL model. Our analysis reveals that three exoplanets— Qatar-4 b, TOI-1278 b, and WASP-173 A b —exhibit detectable radio signals suitable for observation with the SKA telescope.
%\tiny
%All article types: you may provide up to 8 keywords; at least 5 are mandatory.
\end{abstract}

%\Keywords{Planetary radio emissions, exoplanet detection, Nancy Roman Telescope, SKA1-low} 
\section{Introduction}
\noindent Observing and measuring the magnetic fields of exoplanets holds paramount importance in contemporary astrophysical research. The magnetic field of a planet serves as a fundamental aspect of its physical makeup, providing crucial insights into its internal structure and dynamical processes. Understanding exoplanetary magnetic fields allows for the investigation of planetary interiors, shedding light on the presence of metallic cores and the mechanisms driving magnetic field generation, such as dynamo processes. Additionally, the interaction between a planet's magnetic field and the surrounding stellar wind profoundly influences its atmospheric dynamics, including the retention and loss of atmospheric gases. This interplay is pivotal in determining a planet's habitability and its potential to support life as we know it. Furthermore, studying exoplanetary magnetic fields offers a glimpse into the diverse array of planetary systems beyond our own, enriching our understanding of planetary formation and evolution across different stellar environments and orbital configurations. \\
\\
%Overall, observational efforts aimed at characterizing exoplanet magnetic fields represent a crucial frontier in astrophysics, promising profound insights into the nature and diversity of planetary systems throughout the universe.\\
%\\
However, the task of observing and measuring the magnetic field of exoplanets presents a significant challenge. The magnetospheres of several solar system bodies, such as Earth, Jupiter, Saturn, Uranus, and Neptune, manifest distinct patterns of nonthermal continuum radiation emission due to their interaction with the solar wind \citep{zarka1998auroral}. The electron cyclotron maser instability (ECMI) mechanism is believed to be the primary source of non-thermal radio emission in planetary magnetospheres, where energetic electrons emit coherent radiation along magnetic field lines \citep{melrose1982electron, dulk1985radio}. Observations of such emissions provide valuable insights into the strength and configuration of exoplanetary magnetic fields. However, despite extensive efforts, direct detection of radio emissions from exoplanets has proven challenging due to the low sensitivity of ground-based radio telescopes \citep{murphy2015limits, lazio2018magnetic, turner2021search}. Recent advancements, particularly in the development of next-generation instruments like the Square Kilometre Array (SKA), offer promising opportunities for more sensitive and comprehensive observations of exoplanetary radio emissions. By detecting and characterizing these emissions, we can gain deeper insights into the magnetic properties, atmospheric dynamics, and potentially the presence of exomoons in exoplanetary systems.
\\
\\
To date, the confirmed count of exoplanets has reached ~5300. In this study, we conduct an assessment of the non-thermal radio emissions originating from these exoplanets, utilizing the RBL model. Subsequently, we identify potential candidates for detection using the Square Kilometre Array (SKA) telescope.

\section{Methodology}

\subsection{Estimation of Exoplanets Radio Emission}

\noindent In this section we follow the methodology outlined in \citep{bagheri2024infraredradiofollowup} to estimate the rdaio emission of exoplanets. Observations of magnetized planets within our Solar System have unveiled a consistent link between the intensity of auroral emissions and the energy flux of the incoming stellar wind \citep{zarka2001magnetically, zarka2004fast, zarka2007plasma, zarka2018Jupiter}. This phenomenon extends across various planetary bodies, including those exhibiting satellite-Jupiter radio emissions \citep{noyola2014detection, noyola2016radio}. The correlation between auroral emissions, stellar wind energy flux, and the strength of the planetary magnetic field is encapsulated within the RBL model (citep{zarka2001magnetically, zarka2004fast, zarka2007plasma}. Mathematically, this model establishes a relationship between the emitted frequency and the magnetic field strength, as outlined by

\begin{equation}
\nu = \frac{eB}{2 \pi m}~,
\end{equation}
where $\nu$ is the emitted frequency, $e$ is the charge of the emitting particles (in this case, electrons), $B$ is the magnetic field strength and $m$ is the mass of the emitting electrons.
\\
\\
Similarly, the radio power from an exoplanet is related to the incident power of the stellar winds as
\begin{equation}
    P_{rad} = 4 \times 10^{18} erg~ s^{-1} \big(\frac{{\dot{M}_{ion}}}{10^{-14} M_\odot ~yr^{-1}}\big)^{0.8} \big(\frac{v_\infty}{400 ~km ~s^{-1}}\big)^{2} \big(\frac{a}{5 ~au}\big)^{-1.6} \big(\frac{\omega}{\omega_J}\big)^{0.8} \big(\frac{M_p}{M_J}\big)^{1.33}~,\label{eqPower}
\end{equation}
where $a$ is the semi-major axis of the planet’s orbit in au, $\dot{M}$ ion is the stellar ionized mass-loss rate, $v_\infty$ is the terminal velocity of the stellar wind, and $\omega$ is the corotation speed \citep{farrell1999possibility}. \\
\\
Terminal velocity of stellar wind can be calculated by
\begin{equation}
     v_{\infty} = 0.75 \times 617.5 \times \sqrt{R_\star/M_\star}~, 
\end{equation}
where $R_\star$ and $M_\star$ are the host star radius and mass \citep{o2018search}. Therefore, the radio flux density ($S_\nu$) emitted by an exoplanet can be estimated using the following equation
\begin{equation}
    S_\nu = \frac{P_{rad}}{\Delta \nu \Omega d^2}~, \label{eqFreq}
\end{equation}
where $P_{rad}$ is the total radio emission power of the exoplanet \eqref{eqPower}, $d$ is the distance between the exoplanet and the observer, and $\nu$ is the frequency of the radio emission,
 \begin{equation}
    \nu_c = 23.5 ~\text{MHz} ~\Big(\frac{\omega}{\omega_J}\Big) \Big(\frac{M_p}{M_J}\Big)^{5/3} \Big(\frac{R_p}{R_J}\Big)^{3}~, \label{eqPlanetfrequency}
\end{equation}
where $R_p$ is the planetary radius. 
Substituting \eqref{eqFreq} into \eqref{eqPower} gives 
\begin{eqnarray}
    S_\nu &=& 7.6~ \text{mJy}~ (\frac{\omega}{\omega_J})^{-0.2} (\frac{M_p}{M_J})^{-0.33} (\frac{R_p}{R_J})^{-3}
    (\frac{\Omega}{1.6 ~ sr})^{-1} (\frac{d}{10~pc})^{-2}
    \\
    &\times& (\frac{a}{1 ~au})^{-1.6}
    (\frac{{\dot{M}_{ion}}}{10^{-11}~M_\odot ~ yr^{-1}})^{0.8} (\frac{v_\infty}{100~ km~s^{-1}})^{2}~,
\end{eqnarray}
where $\omega_J$, $M_J$, and $R_J$, are Jupiter's corotation speed, mass, and radius and $\Omega$ is the beaming solid angle of the emission. In deriving this expression, we have followed \citep{farrell1999possibility} and assumed that the planet will emit ECMI emission between the frequencies $0.3 \nu_c$ and $\nu_c$, where $\nu_c$ is the maximum radiation frequency. \\
\\
The planetary corotation speed ($\omega$) is a fundamental parameter in planetary magnetospheric dynamics. In the context of close-in exoplanets, where tidal locking is prevalent due to intense gravitational forces from the host star, the corotation rate becomes particularly significant. Tidal locking occurs when the orbital period of the planet matches its rotational period, resulting in one hemisphere of the planet permanently facing the star. As a consequence, the corotation rate of tidally locked planets is synchronized with their orbital motion. This synchronization affects the generation and distribution of electromagnetic phenomena within the magnetosphere, such as auroral emissions and field-aligned currents \citep{zarka2001magnetically, seager2002constraining}. To account for tidal locking's influence, we presume that exoplanets within orbits less than $0.1$ au are tidally locked, meaning their orbital period matches their corotation period. For exoplanets beyond this threshold, we employ the Darwin-Radau relation \citep{murray2000solar} 
\begin{equation}
\omega = \sqrt{\frac{f~G~M_p}{R_{eq}^2}\Bigg[\frac{5}{2}(1-1.5~C)^2 + \frac{2}{5}\Bigg]}  
\end{equation}
where $C$ is 0.4 for rocky planets and for gas giant planets $C = 0.25$ and $f$ is the planet's oblateness \citep{hubbard1984planetary}. For the Jovian planet with orbital distance $0.1 > a $ au, we use the oblateness of Jupiter ($f = 0.064$), and for rocky planets, we use the Earth's value which is $f = 0.00335$ \citep{barnes2003measuring}.

\subsection{Detectablity of the Planetary Radio Signal}

\noindent The leading contemporary ground-based telescope for detecting planetary radio emissions is the SKA telescope. Comprising two arrays, SKA1-Low and SKA1-Mid, it covers a broad frequency range from 50 MHz to 14 GHz. With its immense collecting area, totaling one square kilometer when fully operational, the SKA can capture faint radio signals from various cosmic phenomena, including nearby planets and distant galaxies. Its innovative design incorporates advanced signal processing techniques and data analysis algorithms to maximize scientific output while minimizing data volumes and processing requirements. As a collaborative effort involving over 20 countries, the SKA embodies international cooperation and scientific excellence, poised to unveil new frontiers in radio astronomy. For our observations using the SKA, we focus on the SKA1-Low and SKA1-Mid frequency ranges, up to 890 MHz. The imaging sensitivity for SKA1-Low and SKA1-Mid up to 890 MHz is detailed in Table \ref{tab:SKA}, considering a continuum observation with a fractional bandwidth of approximately $\Delta \nu/\nu_c \approx 0.3$, alongside an integration time of 1 hour.
\begin{longtable}{cccc}
\hline\hline
$\nu_{min}$ [MHz]   & $\nu_{mid}$ [MHz]   &$\nu_{max}$ [MHz]    &$\sigma$ [$\mu Jy$]
\\
\hline
50      &60         &69         &163\\
69      &82         &96         &47 \\
96      &114        &132        &26 \\
132     &158        &183        &18 \\
183     &218        &253        &14 \\ 
253     &302        &350        &11 \\
350     &410        &480        &16.8 \\
480     &560        &650        &8.1 \\
650     &770        &890        &4.4 \\
\hline
\caption{Image sensitivity of SKA1-Low and SKA1-mid within the indicated frequency bands for continuum observations \citep{braun2019anticipated}} \label{tab:SKA}
\end{longtable}
\noindent We use the SKA image sensitivity for different ranges of frequencies as summarized in \ref{tab:SKA}. 

\section{Results}
\begin{figure}[ht!]
\centering
\includegraphics[width=75mm]{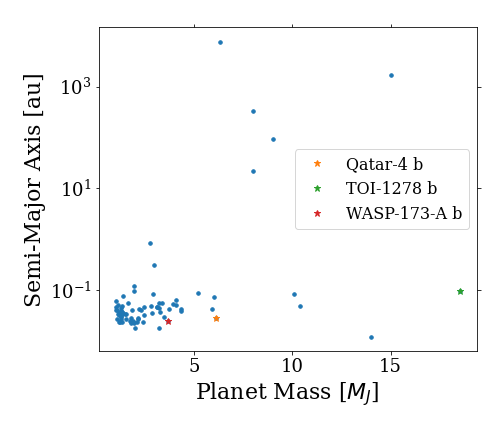}
\includegraphics[width=75mm]{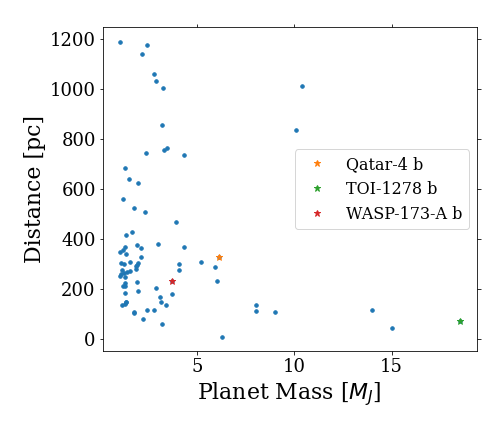}
\\
\includegraphics[width=75mm]{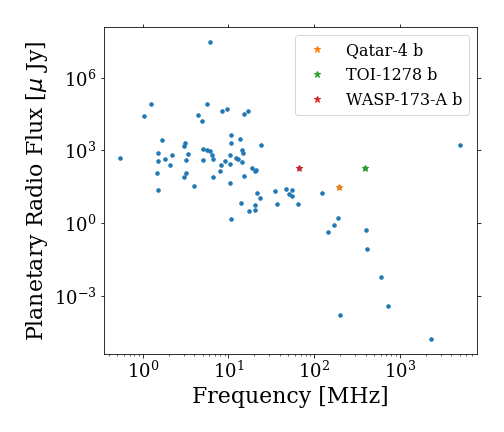}
\includegraphics[width=75mm]{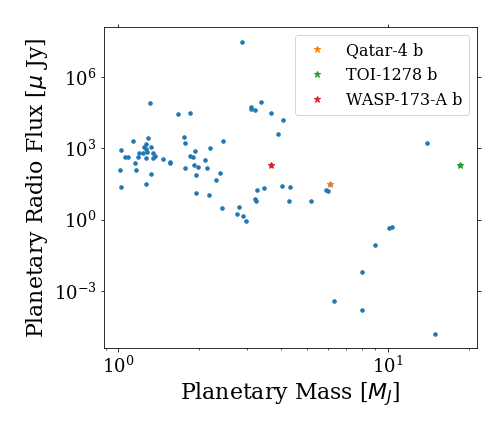}
\caption{Characteristics of 80 confirmed exoplanets. Exoplanets exhibiting detectable radio signals are highlighted in different colors.
\label{fig:scatterPlots}}
\end{figure}

We use the NASA Exoplanet Archive to calculate the radio flux of the confirmed exoplanets. The 80 exoplanets have essential data such as planetary radius and mass, orbital distance, host star radius and mass, and the system's distance from us information. To estimate $\dot{M}_\text{ion}$, we utilized typical mass loss ratios associated with the spectral types of the host stars \citep{michaud2011sirius, puls1996star, krtivcka2014mass, wood2001observational, wargelin2002stringent}. According to our analysis, Exoplanets Qatar-4 b, TOI-1278 b, and WASP-173 A b exhibit radio emissions that are detectable by the SKA telescope. All these three exoplanets are considered as close-in exoplanets. TOI-1278 b is a brown dwarf with an M-type host star. Its mass is 18.5 times that of Jupiter with an orbital period of approximately 14.5 days (0.095 au) \citep{artigau2021toi}. WASP-173 A b is a hot Jupiter ($M_p = 3.69 \times ~ M_J$) orbits a G-type star with an orbital period 1.4 days. \citep{hellier2019new}. Qatar-4 b is also a hot Jupiter exoplanet ($M_p = 6.1 \times ~ M_J$) that orbits a K-type star with an orbital period of 1.8 days \citep{alsubai2017qatar}
\\
\\
Figure \ref{fig:scatterPlots} displays the distributions of planetary parameters, with exoplanets exhibiting detected radio fluxes depicted in distinct colors. It's noticeable that a significant portion of exoplanets emitting intense radio fluxes do so at frequencies below the sensitivity range of the SKA telescope (below 50 MHz). One method to expand the detection of exoplanets with observable radio signals using current technology involves leveraging microlensing events. These events act as natural amplifiers, their magnification unaffected by the radiation's wavelength. A notable characteristic of gravitational lensing is the preservation of radiation polarization during the event. Unlike unpolarized stellar plasma radiation, radio emissions from exoplanets exhibit circular polarization \citep{zarka2014magnetospheric}. Consequently, microlensing events offer a potential means to enhance the detection of exoplanets with observable radio signals \citep{bagheri2024infraredradiofollowup, bagheri2019detection}.

%We repeat our simulation without considering the induced radio emission by exomoons. Figure \ref{fig:noMoons} represents the histograms of detected exoplanet parameters. As seen, the efficiency is much less than that of exomoons. 

\section{Discussion}

\noindent Our understanding of planetary magnetic fields is predominantly derived from observations within our solar system. Diversifying our investigations and insights into exoplanetary magnetic fields could offer valuable perspectives on their internal dynamics and compositions.\\ 
\\
In this study, employing the RBL model, we have identified three exoplanets—Qatar-4 b, TOI-1278 b, and WASP-173 A b—as potential candidates for detecting radio signals using the forthcoming SKA telescope. These exoplanets are characterized as hot Jupiters. Although ECMI can be generated in hot Jupiter magnetospheres, the escape of these radio emissions from these exoplanets may be hindered by their ionospheres \citep{koskinen2013escape, weber2017expanded}. The generation of ECMI requires a specific balance between plasma and cyclotron frequencies, with a depleted and highly magnetized plasma being essential for emission \citep{griessmeier2007predicting, zarka2018Jupiter}. Close-in exoplanets experience significant atmospheric expansion due to factors like tidal forces and XUV-driven heating, leading to the formation of gaseous envelopes around the planets \citep{shaikhislamov2020three, johnstone2018upper, johnstone2019extreme}. Studies suggest that the high plasma densities in these expanded atmospheres may prevent the escape of radio emissions or inhibit the generation of radio waves via ECMI \citep{weber2017expanded}. Therefore, detecting radio emissions from even one hot Jupiter would offer valuable insights into planetary responses to their environments and evolutionary paths, shedding light on the intricate interplay between magnetic fields, stellar wind, and atmospheric escape mechanisms. This knowledge would contribute significantly to our understanding of planet formation, migration, and potential habitability beyond our solar system.
\\
\\
Understanding the magnetic fields and magnetospheric emissions of exoplanets represents a cutting-edge scientific frontier for the coming decade, as emphasized in the Origins, Worlds, and Life Planetary Science \& Astrobiology Decadal Survey report. This report identifies two Priority Science Question Topics that encompass aspects of planetary magnetic fields and their interaction with the solar wind. These topics are labeled as Q6: "Solid Body Atmospheres, Exospheres, Magnetospheres, and Climate Evolution" and Q12.7: "Exoplanets, Giant Planet Structure and Eevolution" \citep{national2022origins}..
\\
\\
%This will open a new window to exoplanetary research.
%The proposed research plan can pave the way toward the first successful detection of an exomoon.  \\
%\\
%Examples include Jovian planets around the young stars [Wood et al., 2002, 2005] or even indirect detection of exomoons via their effects on their parent exoplanets
\section{acknowledgements}
\noindent We acknowledge the support of the US National Science Foundation (NSF) under Grant No. 2138122.
%\end{acknowledgements}

\bibliographystyle{aasjournal} %  Many Frontiers journals use the Harvard referencing system (Author-date), to find the style and resources for the journal you are submitting to: https://zendesk.frontiersin.org/hc/en-us/articles/360017860337-Frontiers-Reference-Styles-by-Journal. For Humanities and Social Sciences articles please include page numbers in the in-text citations
\bibliography{manuscript}

%%% Make sure to upload the bib file along with the tex file and PDF
%%% Please see the test.bib file for some examples of references

%%% Please be aware that for original research articles we only permit a combined number of 15 figures and tables, one figure with multiple subfigures will count as only one figure.
%%% Use this if adding the figures directly in the manuscript, if so, please remember to also upload the files when submitting your article
%%% There is no need for adding the file termination, as long as you indicate where the file is saved. In the examples below the files (logo1.eps and logos.eps) are in the Frontiers LaTeX folder
%%% If using *.tif files convert them to .jpg or .png
%%%  NB logo1.eps is required in the path in order to correctly compile front page header %%%

%%% If you are submitting a figure with subfigures please combine these into one image file with part labels integrated.
%%% If you don't add the figures in the LaTeX files, please upload them when submitting the article.
%%% Frontiers will add the figures at the end of the provisional pdf automatically
%%% The use of LaTeX coding to draw Diagrams/Figures/Structures should be avoided. They should be external callouts including graphics.
%\newpage
%\section*{Appendix A: CCMC simulations Used in This Study}
%\appendix
%\input{table.tex}
%\includepdf[pages={1-32}]{JH_appendix.pdf}

\end{document}